\documentclass[twocolumn]{jpsj3}
\usepackage{bm,amssymb,amsmath,xspace}
\usepackage{times}
\usepackage{color}
\usepackage{ulem}
\newcommand{\simg}{\stackrel{>}{_\sim}}

\newcommand*\wtow    {\textsc{wien2wannier}\xspace}
\newcommand*\wannier {\textsc{wannier90}\xspace}

\title{Fermi Surface, Pressure-Induced Antiferromagnetic Order, and Superconductivity in FeSe}

\author{Jun Ishizuka$^1$, Takemi Yamada$^2$, Yuki Yanagi$^3$, and Yoshiaki \=Ono$^4$ 
}
\inst{$^1$Department of Physics, Kyoto University, Kyoto 606-8502, Japan \\
$^2$Department of Physics, Faculty of Science and Technology, Tokyo University of Science, Noda 278-8510, Japan \\
$^3$Department of Physics, School of Science and Technology, Meiji University, Kawasaki 214-8571, Japan\\
$^4$Department of Physics, Niigata University, Ikarashi, Niigata 950-2181, Japan
}
\abst{
The pressure dependence of the structural ($T_s$), antiferromagnetic ($T_m$), and superconducting ($T_c$) transition temperatures in FeSe is investigated on the basis of the 16-band $d$-$p$ model. At ambient pressure, a shallow hole pocket disappears due to the correlation effect, as observed in the angular-resolved photoemission spectroscopy (ARPES) and quantum oscillation (QO) experiments, resulting in the suppression of the antiferromagnetic order, in contrast to the other iron pnictides. The orbital-polarization interaction between the Fe $d$ orbital and Se $p$ orbital is found to drive the ferro-orbital order responsible for the structural transition without accompanying the antiferromagnetic order. The pressure dependence of the Fermi surfaces is derived from the first-principles calculation and is found to well account for the opposite pressure dependences of $T_s$ and $T_m$, around which the enhanced orbital and magnetic fluctuations cause the double-dome structure of the eigenvalue $\lambda$ in the Eliashberg equation, as consistent with that of $T_c$ in FeSe.
}

\begin{document}
\maketitle

\section{Introduction}
\label{intro}

The discovery of iron-based superconductors \cite{kamihara} is one of the highlights of condensed matter physics and offers us a means of understanding fundamental phenomena of electronic properties. In most iron-based superconductors, the superconductivity is found in the proximity of a magnetic ordered state. Just above the magnetic ordered state (in the 1111 systems such as LaFeAsO and 122 systems such as BaFe$_2$As$_2$), a structural (nematic) transition from the tetragonal to the orthorhombic phase is realized that spontaneously breaks the fourfold symmetry $C_4$. The relation between the structural transition and the magnetic ordered phase has been a controversial issue, the understanding of which may provide insights into the pairing mechanism and symmetry \cite{mazin,kuroki,ikeda,thomale,wang,kontani_orbiton,yanagi_e-ph_2}. To elucidate the origin of the structural transition above the magnetic order, spin-nematic theory \cite{fernandes,fernandes_2} and orbital order theory \cite{kruger,onari_sc-vc} have been intensively proposed. The microscopic pictures of these theories are the spin-quadrupole order induced by the spin fluctuation \cite{fernandes} and the orbital-spin mode-coupling characterized by the Aslamazov--Larkin vertex correction \cite{onari_sc-vc}, respectively. On the other hand, another candidate mechanism of the orbital order was suggested in Ref.~\citenum{yamada_upd} by taking into account an orbital-polarization interaction, which was derived from the orbital dependence of the intersite $d$-$p$ Coulomb integrals between Fe $d_{zx/yz}$ orbitals and As $p_{x/y}$ orbitals. 

The surprising phase diagram of FeSe has attracted much attention in investigating the origin of the superconductivity and nematicity in iron-based superconductors. This material shows a structural transition at $T_s\sim90$ K and a superconducting transition at $T_c\sim9$ K without long-range magnetic order. $T_s$ ($T_c$) is decreased (increased) by isovalent doping in Fe(Se$_x$S$_{1-x}$) \cite{mizuguchi_3,watson_2} and Fe(Se$_x$Te$_{1-x}$) \cite{kawasaki}, and by applying pressure \cite{medvedev}, whereas $T_s$ and $T_c$ are suppressed by nonmagnetic impurity (Co) doping \cite{urata}. Toward $T_s$, the softening of the elastic constant $C_{66}$ \cite{zvyagina} and the enhancement of Raman nematic susceptibility $\chi^{c}_{x^2-y^2}$ \cite{massat} are observed. These results clearly indicate the existence of ferro-orbital fluctuation and electronic ferro-orbital order, as is recognized in other iron-based superconductors, both theoretically \cite{fernandes,onari_sc-vc,yanagi_e-ph_2,yamada_upd} and experimentally \cite{goto,yoshizawa,gallais}. The most remarkable feature of FeSe is observed in the temperature dependence of the spin-lattice relaxation rate $1/T_1T$ in NMR experiments \cite{baek_3,bohmer}. With decreasing temperature, $1/T_1T$ decreases and reaches a minimum at $T\sim T_s$. With further decreasing temperature, $1/T_1T$ starts to increase toward the superconducting temperature.
The temperature dependence of $1/T_1T$ together with NMR Knight shift is interpreted by the low-energy properties of the effective tight-binding model derived in Ref.~\citenum{mukherjee} using a priori information of recent angular-resolved photoemission spectroscopy (ARPES) and quantum oscillation (QO) measurements.

Above $T_s$ (at a high temperature), ARPES experiments \cite{maletz,watson,shimojima_str_2,suzuki} found that hole Fermi surfaces consist of two small pockets of mainly $d_{zx}$ and $d_{yz}$ character around the $\Gamma$-$Z$ line. A narrow $d_{xy}$ hole band also exists in $\sim$50 meV below the Fermi level. The orbital-dependent mass enhancements were estimated as $d_{zx/yz}\sim 3$ and $d_{xy}\sim 8$. ARPES experiments also found a small electron Fermi surface at the $M$ point of mainly $d_{zx}$ and $d_{yz}$ characters. The QO measurement \cite{watson} performed at a low temperature detected the presence of the $d_{xy}$ electron pocket, which is difficult to observe in ARPES. Below $T_s$, ARPES observed large band splitting of $\sim50$ meV at the $M$ point, which corresponds to the orbital order breaking the degeneracy of the $d_{zx}$ and $d_{yz}$ orbitals. The orbital order has momentum-dependent sign inversion \cite{suzuki} as $E_{yz}(\Gamma)-E_{zx}(\Gamma)\sim-10$ meV and $E_{yz}(M)-E_{zx}(M)\sim+50$ meV, which is interpreted on the basis of the orbital order scenario \cite{onari_FeSe,yamakawa_FeSe_sc-vc,yamakawa_FeSe_p} of Aslamazov--Larkin vertex corrections.

FeSe gives us a remarkable phase diagram as a function of pressure. An early powder sample study \cite{medvedev} showed that $T_c$ can be enhanced by $\sim37$ K at a pressure $\sim9$ GPa. Polycrystalline samples of FeSe$_{1-x}$ were investigated in early NMR measurements and muon-spin rotation ($\mu$SR) experiments \cite{imai,bendele,bendele_2}. According to these results, the spin fluctuation is enhanced at pressures up to $\sim$2.2 GPa \cite{imai} and a pressure-induced antiferromagnetic ordered state is stabilized above $\sim$1 GPa \cite{bendele,bendele_2}. More recent studies on single crystals revealed the comprehensive and complex $T$-$P$ phase diagram from macro- and microscopic measurements under high pressure \cite{terashima,sun,miyoshi_2,kothapalli,terashima_2,wang_6}. 
Remarkably, a nonmagnetic structural transition $T_{s}$ is quickly suppressed at pressures up to $P\sim1$ GPa \cite{miyoshi_2} and the antiferromagnetic ordered state is stabilized in a wide pressure region ($1<P<6$ GPa) with a dome shape at the maximum value of $T_m\sim45$ K ($P\sim4.8$ GPa) \cite{sun}. Thus, the superconductivity meets three quantum critical points at $T=0$ with increasing $P$. One is the nematic quantum critical point, the others are magnetic ones accompanied by the enhancement of $T_c$. It is noteworthy that the stripe-type antiferromagnetic order breaks not only time-reversal symmetry but also $C_4$ symmetry. Simultaneous first-order magnetic and structural transitions were observed in transport measurement \cite{kothapalli}.

To elucidate the electronic correlation effects on FeSe, theoretical studies have been intensively performed \cite{liebsch_2,aichhorn_3,yin_power-low,miyake,hirayama,scherer}. The constrained random phase approximation (cRPA) \cite{miyake} combined with the {\it ab initio} calculation scheme indicates that the correlation strength plays an important role in understanding the material dependence of the iron-based superconductors; for instance, the Coulomb interaction of FeSe is larger than that of LaFeAsO. The larger Coulomb interaction yields larger mass enhancement with significant orbital dependence \cite{aichhorn_3,yin_power-low}. However, the origin of unusual electronic states observed in experiments \cite{maletz,watson} has been a controversial issue.

In contrast to the 1111 and 122 systems, FeSe at ambient pressure shows a very weak antiferromagnetic fluctuation above $T_s$. The absence of the low-energy spin response seems to be consistent with orbital-polarization interaction mechanism \cite{yamada_upd}, which merely relies on the orbital degrees of freedom in principle, as mentioned above.
Furthermore, the average value of the intersite Coulomb integrals is large for FeSe relative to those for the 1111 and the systems \cite{miyake}. Thus, by considering suitable intersite $d$-$p$ Coulomb integrals, FeSe is expected to have larger orbital-polarization interaction than the 1111 and 122 systems.

In this article, we show a detailed analysis of a 16-band $d$-$p$ model for FeSe. In Sect.~\ref{sect2}, we show that the spectral function of the $xy$ orbital derived from dynamical mean-field theory (DMFT) deviates from the first-principles calculation owing to the orbital dependence of the 3$d$ Coulomb interaction. The absence of a hole Fermi surface composed of the $xy$ orbital is in agreement with experiments \cite{maletz,watson}. The experimental results of the NMR spin-relaxation rate \cite{baek_3,bohmer}, in which the spin fluctuation is very weak above the structural transition, was verified. We show that the orbital fluctuation enhanced owing to the orbital-polarization interaction \cite{yamada_upd}. The enhancement explains the elastic constants in ultrasonic experiments \cite{zvyagina} and the susceptibility in Raman scattering experiments \cite{massat}. 
In Sect.~\ref{sect3}, we present the pressure dependence of the spin and orbital fluctuations and the superconductivity using the RPA calculation. 
Here, we assume the derived Coulomb interaction parameters at ambient pressure and neglect the correlation effect for simplicity.
The pressure dependence of the electronic state was derived from first principles.
From the first-principles analysis, the temperature-pressure phase diagram of FeSe can be naturally understood. The orbital-dependent Coulomb interaction not only reproduces the key properties of FeSe well but also provides a helpful perspective to investigate the strongly correlated multiorbital systems.


\section{Electronic State at Ambient Pressure}
\label{sect2}
In this section, we use the model of FeSe at ambient pressure derived from first principles and solve this model by DMFT to investigate the effect of the orbital dependence of the Coulomb interaction. We show how the $d_{xy}$ hole band lies under the Fermi level predicted by several experiments and elucidate the origin of  the orbital-energy shift.
Consequently, the spin fluctuation due to the $xy$ orbital is clearly suppressed. Moreover, we show the orbital fluctuation enhancement by introducing an ``orbital-polarization $d$-$p$ Coulomb interaction'' \cite{yamada_upd}, which is not attributable to the spin degrees of freedom.
\subsection{Dynamical mean field theory}
\label{sect2-1}
In this subsection, we illustrate the DMFT \cite{ishizuka_s+-s++,ishizuka_hole-s+-} for the multiorbital $d$-$p$ model on the FeSe system. 

First, we introduce a 16-band $d$-$p$ model including the multiorbital Coulomb interactions $H_{dd}$ between 3$d$ electrons on the Fe site and $H_{pp}$ between 4$p$ electrons on Se site and the intersite Coulomb interaction $H_{dp}$ between 3$d$ and 4$p$ electrons,
\begin{eqnarray}
H=H_0+H_{dd}+H_{pp}+H_{dp}.
\end{eqnarray}
$H_0$ is the tight-binding Hamiltonian derived from first principles
using maximally localized Wannier functions \cite{marzari,souza,mostofi,kunes}. The electronic structure is described by density functional theory using the {\textsc{wien}2k\xspace} code \cite{blaha_2,wien2k}. Diagonalizing $H_0$, we obtain the non-interacting band structure and the Fermi surface. 

The on-site Coulomb interaction part on the Fe site is given as
\begin{eqnarray}
H_{dd}&=&\sum_{i}\sum_{l}
U_{ll}d^{\dag}_{il\uparrow}d^{\dag}_{il\downarrow}
d_{il\downarrow}d_{il\uparrow} \nonumber \\
&+&\sum_{i}\sum_{l>l'}\sum_{\sigma,\sigma'}
U_{ll'}d^{\dag}_{il\sigma}d^{\dag}_{il'\sigma'}
d_{il'\sigma'}d_{il\sigma} \nonumber \\
&+&\sum_{i}\sum_{l>l'}\sum_{\sigma,\sigma'}
J_{ll'}d^{\dag}_{il\sigma}d^{\dag}_{il'\sigma'}
d_{il\sigma'}d_{il'\sigma} \nonumber \\
&+&\sum_{i}\sum_{l>l'}\sum_{\sigma\neq\sigma'}
J_{ll'}d^{\dag}_{il\sigma}d^{\dag}_{il\sigma'}
d_{il'\sigma'}d_{il'\sigma}, \label{eq_H_dd}
\end{eqnarray}
where $d^{(\dag)}_{il\sigma}$ is the annihilation (creation) operator of the 3$d$ electron with orbital $l$ spin $\sigma$ on site $i$, and $U_{ll}, U_{ll'}, J_{ll'}, J_{ll'}$ are the Coulomb interaction matrix of the intra- and inter-orbital direct terms, the Hund's rule coupling, and the pair transfer, respectively. From the first-principles downfolding scheme given by the cRPA method, Ref.~\citenum{miyake} revealed that $U_{ll}, U_{ll'}, J_{ll'}, J_{ll'}$ in $H_{dd}$ are orbital-dependent and that the average of $U_{ll}$ is ${\bar U}_d=7.2$ eV for FeSe.

The on-site Coulomb interaction part on Se site is given as
\begin{eqnarray}
H_{pp}&=&\sum_{i}\sum_{m}
U_{mm}p^{\dag}_{im\uparrow}p^{\dag}_{im\downarrow}
p_{im\downarrow}p_{im\uparrow} \nonumber \\
&+&\sum_{i}\sum_{m>m'}\sum_{\sigma,\sigma'}
U_{mm'}p^{\dag}_{im\sigma}p^{\dag}_{im'\sigma'}
p_{im'\sigma'}p_{im\sigma} \nonumber \\
&+&\sum_{i}\sum_{m>m'}\sum_{\sigma,\sigma'}
J_{mm'}p^{\dag}_{im\sigma}p^{\dag}_{im'\sigma'}
p_{im\sigma'}p_{im'\sigma} \nonumber \\
&+&\sum_{i}\sum_{m>m'}\sum_{\sigma\neq\sigma'}
J_{mm'}p^{\dag}_{im\sigma}p^{\dag}_{im\sigma'}
p_{im'\sigma'}p_{im'\sigma}, \label{eq_H_pp}
\end{eqnarray}
where $p^{(\dag)}_{im\sigma}$ is the annihilation (creation) operator of the 4$p$ electron with orbital $m$ spin $\sigma$ on site $i$, and $U_{mm}, U_{mm'}, J_{mm'}$ are the Coulomb interaction matrix. ${\bar U}_p=4.7$ eV was estimated for FeSe from cRPA study \cite{miyake}.
We assume $(U_{mm}, U_{mm'}, J_{mm'})=(U_p, U'_p, J_p)$, $U_p=U'_p+2J_p$, and $J_p/U_p=0.1$.
Hereafter, labels $l$ and $m$ represent Fe-3$d$ and Se-4$p$ orbital indices, respectively.

The intersite Coulomb interaction part is written as
\begin{equation}
H_{dp}=V\sum_{\langle i,j\rangle}n_{di}n_{pj}+V'\sum_{\langle i,j\rangle}(n_{dizx}-n_{diyz})(n_{pjx}-n_{pjy}),
\label{eq_H_dp}
\end{equation}
where $n_{dil}$ ($n_{pjm}$) is the number operator of a $d$ ($p$) electron with orbital $l$ ($m$) on site $i$ ($j$), $n_{di}=\sum_ln_{dil}$ ($n_{pj}=\sum_mn_{pjm}$), and $\langle i,j\rangle$ represents the summation of nearest-neighbor Fe and Se sites. In Eq.~(\ref{eq_H_dp}), $V'$ is the $d$-$p$ orbital-polarization interaction, which has been found to enhance the orbital fluctuation as discussed for the iron-pnictides \cite{yamada_upd}, originating from the orbital dependence of the Coulomb integrals $V_{lm}$ between Fe $d$ orbitals and Se $p$ orbitals: $V'=(V_{zx,x}-V_{zx,y})/2$.

In DMFT with a sublattice degree of freedom, the original site is mapped onto the effective impurity system for each sublattice. Hence, one needs to solve the impurity problem twice on two adjacent sites of the original lattice. Note that the $d$ orbital for the five- or ten-orbital model contains considerable Se $p$ orbital components, and thus the hybridization makes the Wannier function delocalized and anisotropic. This indicates that the $d$-$p$ model is a good starting point of DMFT rather than the $d$ model. In the $d$-$p$ model, since the Coulomb interaction parameters are almost isotropic, the double counting of the correlation effect considered in the local density approximation is expected to be very simple \cite{miyake}. Hence, we argue that the orbital dependence of the double counting is negligibly small. 

The spin (charge-orbital) susceptibility in the $d$-$p$ model is given by
\begin{equation}
\hat{\chi}^{s(c)}(q)=\hat{\chi}^0(q)\left[\hat{1} -(+)\hat{\Gamma}^{s(c)}(q)\hat{\chi}^0(q)\right]^{-1},
\label{eq_chi_chap3}
\end{equation}
where $q=(\bm{q},i\omega_n)$ with the wave vector $\bm q$ and bosonic Matsubara frequency $\omega_n=2n\pi T$. The irreducible susceptibility is defined as $\hat{\chi}_0(q)=-(T/N)\sum_{k}\hat{G}(k+q)\hat{G}(k)$, where $\hat{G}(k)=[(i\varepsilon_m+\mu)-\hat{H}_0(\bm{k})-\hat{\Sigma}(i\varepsilon_m)]^{-1}$ is the lattice Green's function, $\hat{H}_0(\bm{k})$ is the kinetic part of the Hamiltonian with the wave vector $\bm{k}$, $\hat{\Sigma}(i\varepsilon_m)$ is the lattice self-energy, which coincides with the impurity self-energy obtained in the impurity Anderson model, and $k=(\bm{k},i\varepsilon_m=i(2m+1)\pi T)$.
$\hat{\Gamma}^{s(c)}$ is the spin (charge-orbital) vertex, which is given by the following matrix:
\begin{eqnarray}
\hat{\Gamma}^{s}(i\omega_n)=\left[\begin{array}{cc}
\hat{\Gamma}^{s}_{dd}(i\omega_n)\otimes\hat{\sigma}_0& 0 \\
0 & \hat{\Gamma}^{s}_{pp}(i\omega_n)\otimes\hat{\sigma}_0\end{array}\right], \\
\hat{\Gamma}^{c}(q)=\left[\begin{array}{cc}
\hat{\Gamma}^{c}_{dd}(i\omega_n)\otimes\hat{\sigma}_0& \hat{\Gamma}^{c,0}_{dp}(\bm{q}) \\
\hat{\Gamma}^{c,0\ \dag}_{dp}(\bm{q}) & \hat{\Gamma}^{c}_{pp}(i\omega_n)\otimes\hat{\sigma}_0\end{array}\right],
\end{eqnarray}
where $\hat{\sigma}_0$ is the $2\times 2$ identity matrix for the sublattice.
 $\hat{\Gamma}^{s(c)}_{dd(pp)}(i\omega_n)$ is the local irreducible spin (charge-orbital) vertex function in which only the external frequency ($\omega_n$) dependence is considered as a simplified approximation \cite{ishizuka_s+-s++,park} and is explicitly given by
\begin{eqnarray}
\hat{\Gamma}^{s(c)}(i\omega_n)=-(+)\left[\hat{\chi}_{s(c)}^{-1}(i\omega_n)-\hat{\chi}_0^{-1}(i\omega_n)\right],
\label{eq:gamma}
\end{eqnarray} 
with
$
\hat{\chi}_0(i\omega_n)=-T\sum_{\varepsilon_m}
\hat{G}(i\varepsilon_m+i\omega_n)\hat{G}(i\varepsilon_m), 
$
where $\hat{\chi}_{s(c)}(i\omega_n)$ is the local part of the spin (charge-orbital) susceptibility.
The matrix elements in the $d$-$p$ submatrix are
$\Gamma_{llmm}^{c,0}(\bm{q})$=$2(V\pm V')\phi(\bm{q})$,
where $+V'$ for $(l,m)=(zx,x)$ or $(yz,y)$, $-V'$ for $(l,m)=(zx,y)$ or $(yz,x)$, and otherwise $V'=0$.
$\phi(\bm{q})=\sum_{\langle i,j\rangle}e^{i{\bm q}({\bm R}_i-{\bm R}_j)}$ represents the $\bm{q}$-dependent factor due to intersite Fe-Se contributions, where ${\bm R}_i-{\bm R}_j$ denotes the lattice vector.
Here, we ignored the ladder-type Feynman diagrams for $V'$ in $\hat{\Gamma}^{c,0}(\bm q)$.

%

The spin susceptibility obtained from DMFT is reduced relative to that from the RPA due to the local correlation effect but is still overestimated as the non local correlation is neglected. In order to avoid the divergence of spin susceptibility, we introduce a reduction factor $f_{d}$ or $f_p$, that is, $f_dH_{dd}$ or $f_pH_{pp}$ and $f_{d(p)}<1$, while keeping the relative orbital dependence of the Coulomb interaction, as previously done in the RPA study \cite{yamada_upd}. Hereafter, we set $f_d\sim0.5$ for DMFT, $f_d\sim0.25$ for the RPA, $f_d=f_p$. For the intersite $d$-$p$ Coulomb interaction, $V=0.4$ eV and $V'$ are assumed to be parameters including the reduction factors and are set to be smaller than the cRPA value of $1.7$ eV given in Ref.~\citenum{miyake}.
We use the exact diagonalization method for a finite-size cluster as an impurity solver to obtain the local quantities such as the self-energy, which is calculated at $T=0$ as the explicit $T$-dependence is expected to be small in the intermediate correlation regime with $Z\simg0.5$. For the double-counting  correction, we use the fully localized limit (FLL) formula \cite{anisimov_fll}.

The implementation of DMFT in the {\it ab initio} downfolding model with the $d$ and $p$ orbital Coulomb interaction is firstly reported in the present study. Although the double-counting problem is at present a major difficulty, the application of the FLL formula with only the $d$ orbital interaction gives similar results for $f_d=0.5$, except for a small difference in the $d$-$p$ hybridization gap between the conduction bands and valence bands. Thus, we expect that the present calculation is sufficiently accurate at least up to the intermediate regime.

\subsection{Results}
\label{sect2-2}

\begin{figure*}[t]
\begin{center}
\includegraphics[width=170mm]{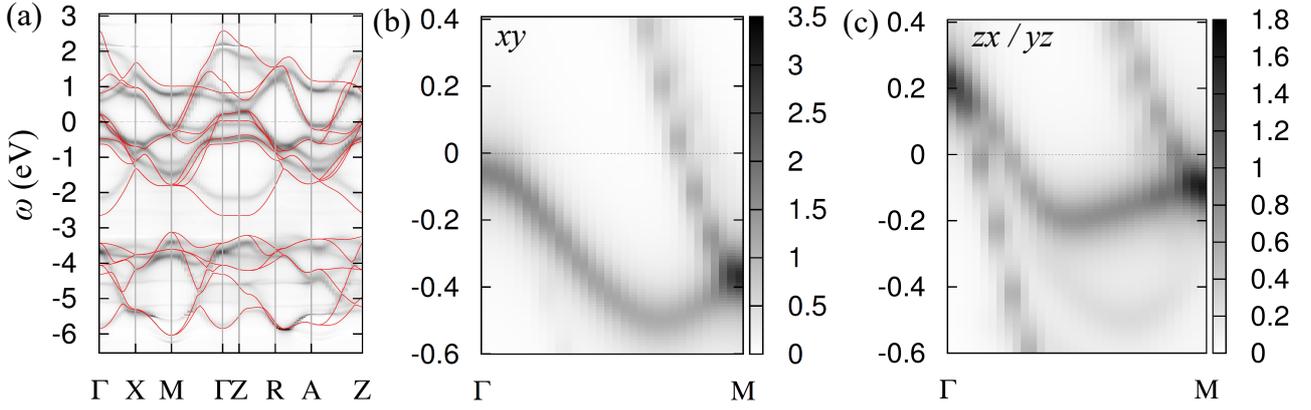}
\caption{(Color online)
 Spectral function $A({\bm k},\omega)$ for orbital-averaged components (a), $xy$ orbital components (b), and $zx/yz$ orbital components (c). The non-interacting band structure is depicted as red solid lines.
\label{fig_a_chap3}}
\end{center}
\end{figure*}

We respectively show in Figs.~\ref{fig_a_chap3}(a)-\ref{fig_a_chap3}(c) the orbital-averaged, $xy$ orbital, and $zx/yz$ orbital components of the spectral function, which are obtained by numerical analytic continuation to $i\omega_n\rightarrow\omega+i\delta$ using the Pad$\acute{\rm e}$ approximation, along high-symmetry directions with $T=0.03$ eV and $f_{d}=f_{p}=0.5$, that is, $U_{xy}=3.55$ eV, $U_{zx/yz}=3.63$ eV, and $U_p=2.35$ eV. Note that the renormalization factor defined by $Z_{l}=\left[1-\frac{d\Sigma_{l}(\varepsilon)}{d(\varepsilon)}\bigl.\bigr|_{\varepsilon\rightarrow0}\right]^{-1}$ is $Z_{xy}=0.77$ for $f_d=0.5$ and $Z_{xy}=0.16$ for $f_d=1$, as is predicted in DMFT with the continuous-time quantum Monte Carlo \cite{aichhorn_3}. The orbital-averaged spectral function exhibits overall renormalizations of quasiparticle bands, even in Se $p$ bands. The orbital-resolved spectral function shows that a hole Fermi surface of $xy$ orbital character around the $\Gamma$ point is pushed downward below the Fermi level $E_{xy}(\Gamma)\sim-50$ meV. This is because the deviation of the Coulomb repulsion yields the orbital dependence of the self-energy in the real part, resulting in the crystal field splitting owing to the electronic correlation. The obtained electronic state captures the experimental result, namely the absence of the $xy$-orbital hole pocket \cite{maletz,watson}.

\begin{figure}[t]
\begin{center}
\includegraphics[width=82mm]{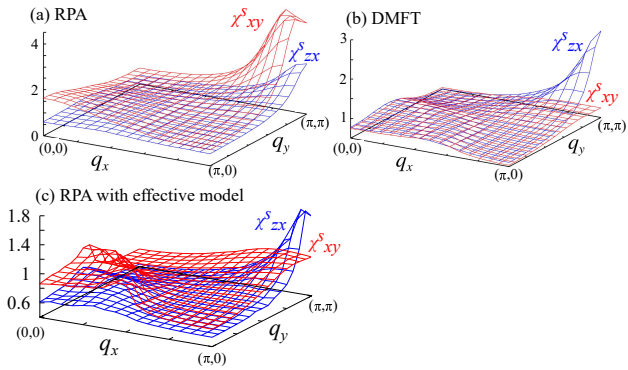}
\caption{(Color online)
 Orbital-resolved spin susceptibility in $q_x$-$q_y$ plane for the RPA (a) and (c), and DMFT (b) at $q_z=0$ and $i\omega_n=0$. Here, $f_d=0.3$ for the RPA, and $f_d=0.5$ for DMFT. In the calculation of (c), the $xy$ orbital energy level is shifted by $-0.14$  eV. 
\label{fig_chis_chap3}}
\end{center}
\end{figure}

In Figs.~\ref{fig_chis_chap3}(a) and \ref{fig_chis_chap3}(b), we show the $xy$ and $zx$ orbital components of the spin susceptibility obtained by the RPA and DMFT, respectively. Because of the absence of electron-hole nesting in the $xy$ orbital, the relation $\chi^s_{xy}\ll\chi^s_{zx}$ is found in DMFT, whereas $\chi^s_{xy}\gg\chi^s_{zx}$ in the RPA. It is recognized that weak spin susceptibility and $\chi^s_{xy}\ll\chi^s_{zx}$ are satisfied within the RPA \cite{mukherjee} when the {\it ab initio} tight-binding model is adjusted to reproduce ARPES and QO results, and the spin-lattice relaxation rate is weak above $T_s$, consistent with experiments \cite{baek_3,bohmer}. From our results, the weakness of the spin fluctuation is confirmed from a self-energy correction.

\begin{figure}[t]
\begin{center}
\includegraphics[width=82mm]{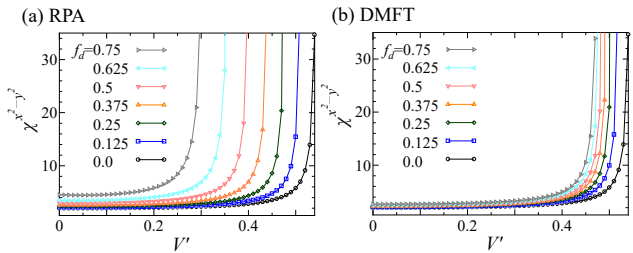}
\caption{(Color online)
 Ferro-orbital (nematic) susceptibility for RPA (a) and DMFT (b) at ${\bm q}=(0,0,0)$ and $i\omega_n=0$. In DMFT, the non local self-energy is omitted as mentioned in the text. The RPA data for $f_d=0$ coincides that for DMFT.
\label{fig_chiquad_chap3}}
\end{center}
\end{figure}

Next, we address the effect of the intersite Coulomb interaction. In our DMFT, the intersite interaction is considered by RPA, since the intersite self-energy correlation is expected to be negligibly small except in the proximity of the critical point. We show in Fig.~\ref{fig_chiquad_chap3} that the ferro-orbital (nematic) susceptibility defined as $\chi^{c}_{x^2-y^2}(\bm{0},0)=\sum_{l_1,l_2,l_3,l_4}o^{l_1l_2}_{x^2-y^2}\chi^{c}_{l_1,l_2;l_3,l_4}(\bm{0},0)o^{l_4l_3}_{x^2-y^2}$ with $o^{zx,zx}_{x^2-y^2}=-o^{yz,yz}_{x^2-y^2}=-(\sqrt{3}/2)o^{3z^2-r^2,x^2-y^2}_{x^2-y^2}=1$ \cite{kontani_orbiton} is enhanced by the orbital-polarization interaction $V'$ for $f_d=0$. 
In addition, the on-site Coulomb interaction ($f_d$) also enhances the orbital susceptibility for both the DMFT and RPA cases as shown in Fig.~\ref{fig_chiquad_chap3}. Similar to the spin susceptibility, the orbital susceptibility from DMFT is reduced relative to that from the RPA, especially for large $f_{d}$ due to the local correction effect. For a typical value $f_{d}=0.5$, the critical interaction for the orbital (nematic) order is $V'\sim0.48$ eV (DMFT) and $V'\sim0.4$ eV (RPA).
Note that the Hund's coupling always suppresses the orbital fluctuation. The ratio between the Hund's coupling and Coulomb interaction is $\bar{J}_d/\bar{U}_d=0.0945$ in FeSe and $\bar{J}_d/\bar{U}_d=0.134$ in LaFeAsO. The smallness of $\bar{J}_d/\bar{U}_d$ results in the $\chi^{c}$ enhancement in FeSe being larger than that of LaFeAsO. 


\section{Electronic State Under Pressure}
\label{sect3}

In this section, we study the pressure dependence of the spin and orbital fluctuations in paramagnetic FeSe within the RPA instead of DMFT because of the following two reasons: (1) the DMFT calculation for each pressure requires considerable CPU time, and (2) the spin susceptibility from DMFT is well reproduced by that from the RPA with the use of the low-energy effective model (see later) as shown in Fig.~\ref{fig_chis_chap3}(c).

\subsection{Construction of the $d$-$p$ model}
\label{sect3-1}

To accurately predict the electronic structure from first principles, we employ a downfolding scheme in the following. We construct the global band structure using density functional theory within the generalized gradient approximation \cite{blaha_2,wien2k}. To minimize the lattice energy from first principles, we start with the structural optimization of the internal coordinate $h_{\rm Se}$ and $c/a$ ratio in each pressure with the space group $P4/nmm$ and the experimental crystal parameters of FeSe \cite{medvedev,margadonna}. We have optimized the structure requiring the Se atomic force to be less than 0.5 mRy/bohr. The optimized lattice parameter $a$ is 2$\%$ smaller (and $c$ is 2$\%$ larger) than those in experiments at ambient pressure. 
%

We construct a tight-binding Hamiltonian exploiting the maximally localized Wannier functions including Fe-3$d$ orbitals and Se-4$p$ orbitals using the \wannier code \cite{mostofi} through the \wtow interface \cite{kunes}.
We introduce the intraorbital hopping parameters $(\delta \varepsilon_{xy},\delta t^{\rm nn}_{xy},\delta t^{\rm nnn}_{xy})=(-0.025,+0.0125,-0.06625)$ [eV] into the tight-binding Hamiltonian, which corresponds to the energy shift of the $xy$ orbital band of $(\delta E_\Gamma, \delta E_{\rm M}, \delta E_{\rm X})=(-0.24,-0.34,+0.24)$ [eV] for the unfolded Brillouin zone. We also introduce the intraorbital hopping parameters $(\delta \varepsilon_{zx/yz},\delta t^{\rm nn}_{zx/yz},\delta t^{\rm nnn}_{zx/yz})=(0,-0.03,-0.03)$ [eV], where the energy shift of the $zx/yz$ orbital band is $(\delta E_\Gamma, \delta E_{\rm M}, \delta E_{\rm X})=(-0.24,0,+0.12)$ [eV]. These parameters are attributable to the self-energy. Since the experimental Fermi surface under pressure has not been detected clearly, we introduce the same $\delta t_l$ for each pressure as a simplified approximation.

\begin{figure}[ht]
\begin{center}
\includegraphics[width=82mm]{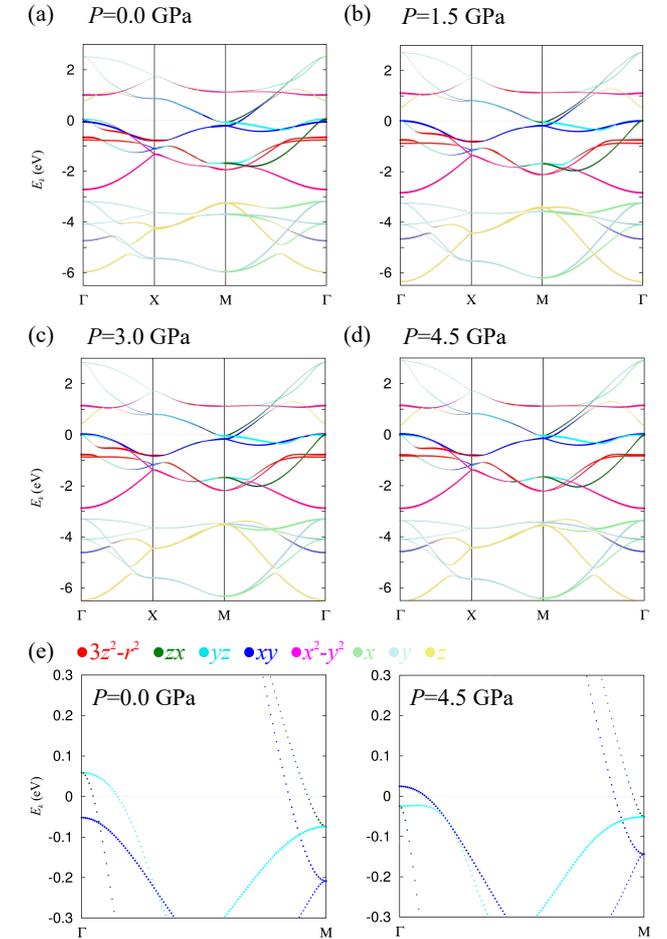}
\caption{(Color online)
 Band structures of FeSe at $P=0.0, 1.5, 3.0,$ and $4.5$ GPa obtained by effective tight-binding models, where orbital weights are plotted as $3z^2-r^2$ (red), $zx$ (green), $yz$ (cyan), $xy$ (blue), $x^2-y^2$ (pink), $x$ (light green), $y$ (light blue), $z$ (gold).
\label{fig_band}}
\end{center}
\end{figure}
\begin{figure}[ht]
\begin{center}
\includegraphics[width=82mm]{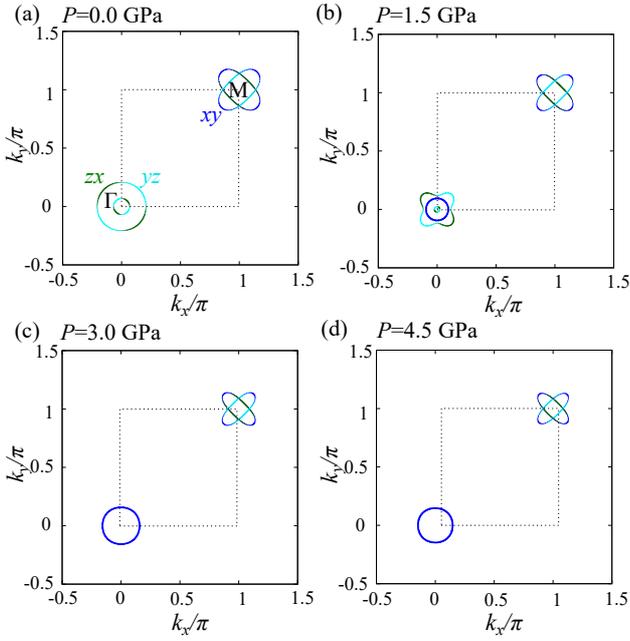}	
\caption{(Color online)
 Fermi surfaces of FeSe at $P=0.0, 1.5, 3.0,$ and $4.5$ GPa obtained by effective tight-binding models, where orbital weights are plotted as $zx$ (green), $yz$ (cyan), $xy$ (blue).
\label{fig_fs}}
\end{center}
\end{figure}

Figure \ref{fig_band} shows the band structures together with the orbital weights at $P=0.0, 1.5, 3.0,$ and $4.5$ GPa. The low-energy band structure for $P=0.0$ GPa is in good agreement with the ARPES results above $T_s$ \cite{watson}. It seems that the band structure of 4.5 GPa resembles that of 0.0 GPa, but total bandwidth is widened by $\sim$1 eV by the pressure effect. Furthermore, by comparing the band structure near the Fermi level at the $\Gamma$ point [see Fig.~\ref{fig_band}(e)], one sees a characteristic orbital energy shift. Namely, the hole band of mainly $xy$ orbital character at $E_\Gamma\sim-50$ meV is pushed upward to $\sim$25 meV and the two degenerated hole bands of mainly $zx/yz$ orbital character at $E_\Gamma\sim60$ meV are pushed downward to $\sim$$-$25 meV, which indicate a topological transition induced by the pressure effect from the first-principles band calculation. It is worth noting that the consequences of the transfer integrals between the $xy$ orbitals together with the Se height are significant for the orbital energy shift \cite{kuroki_4}, which will be discussed later in detail.
To obtain more insight about the low-energy property, the Fermi surfaces at $P=0.0, 1.5, 3.0,$ and $4.5$ GPa with $zx/yz$ and $xy$ orbital weights are shown in Fig.~\ref{fig_fs}. At 0.0 GPa, the electron-hole nesting in the $zx/yz$ orbital is most important. By applying pressure, one can see that the small two-hole pockets deform and shrink, while a new hole pocket of mainly $xy$ orbital character appears. Note that the two-electron pockets slightly change their shape. In the configurations for $P=3.0$ and $4.5$ GPa, the electron-hole nesting in the $xy$ orbital is most important, in contrast to that for $0.0$ GPa.

\begin{figure}[t]
\begin{center}
\includegraphics[width=82mm]{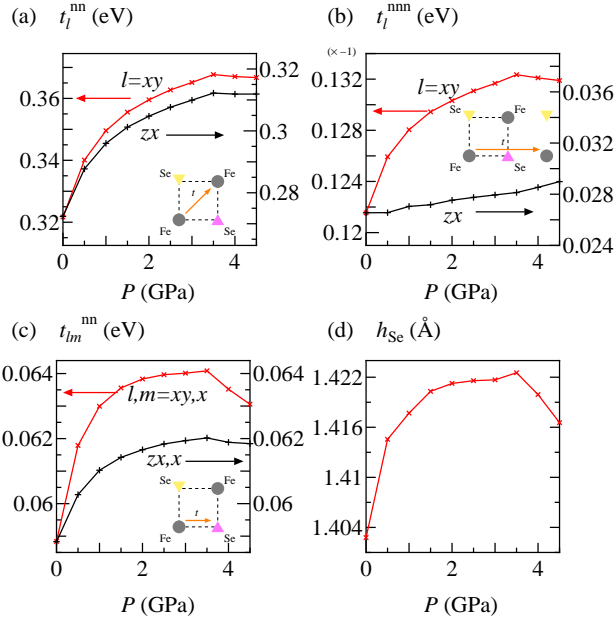}
\caption{(Color online)
 Hopping integrals for (a) intraorbital nearest-neighbor transfer, (b) intraorbital next-nearest-neighbor transfer between Fe atoms, and (c) inter-orbital nearest-neighbor-transfer between Fe and Se atoms with $l=zx,xy$ and $m=x$ as a function of pressure. (d) Se height above the Fe plane.
\label{fig_hopping}}
\end{center}
\end{figure}

Now let us discuss why the hole band of $xy$ orbital character is pushed upward with increasing pressure. This is attributable to the lattice parameter of Se height [see Fig.~\ref{fig_hopping}(d)], which tends to increase with pressure. This effect has been recognized from an early study on the 1111 system, i.e., the lattice parameter of the pnictogen height determines the mixture of the $xy$ orbital contribution, and in turn, the construction of the hole Fermi surface at the $\Gamma$ point \cite{kuroki_4}. To examine this effect on FeSe under pressure, Figs.~\ref{fig_hopping}(a)-\ref{fig_hopping}(c) show the pressure dependences of intraorbital hopping integrals $t_{l}$ and $t_{lm}$ between the lattice vector illustrated in the inset of the Figs.~\ref{fig_hopping}(a)-\ref{fig_hopping}(c) for $(l,m)=(zx/xy,x)$. These hopping integrals tend to more increase with increasing $P$ owing to the decrease in unit-cell volume. The $xy$ hopping integrals have considerably larger pressure dependence than the $zx$ ones. The difference increases as $P$ increases in the low-pressure region, and for further increasing $P$, it starts to decrease for $P>3.5$ GPa.
%

%
%
%

\subsection{Random phase approximation}
\label{sect3-2}

We apply the RPA to each model, where the self-energy correction was neglected. The spin (charge-orbital) susceptibility in the RPA is given by
$\hat{\chi}^{s(c)}(q)=\hat{\chi}^0(q)\left[\hat{1} -(+)\hat{\Gamma}^{s(c),0}(\bm q)\hat{\chi}^0(q)\right]^{-1}$, where $\hat{\Gamma}^{s(c),0}(\bm q)$ is the bare spin (charge-orbital) vertex given in Ref.~\citenum{yamada_upd}.

To examine the superconductivity, we solve the linearized Eliashberg equation,
 \begin{eqnarray}
 \lambda \Delta_{ll'}(k)&=-\frac{T}{N}\sum_{k'}
 \sum_{l_1l_2l_3l_4}V_{ll_1,l_2l'}(k-k') \nonumber \\
 &\times G_{l_3l_1}(-k')\Delta_{l_3l_4}(k') G_{l_4l_2}(k'),
 \label{eq:gap}
 \end{eqnarray}
and obtain the superconducting gap function $\hat{\Delta}(k)$ with the eigenvalue $\lambda$,
which becomes unity at the superconducting transition temperature $T_c$, 
where the effective pairing interaction for the spin-singlet state is given as 
\begin{eqnarray}
\hat{V}(q)&=&
\frac{3}{2}\hat{\Gamma}^{s,0}\hat{\chi}^{s}(q)\hat{\Gamma}^{s,0}
-\frac{1}{2}\hat{\Gamma}^{c,0}(\bm{q})\hat{\chi}^{c}(q)\hat{\Gamma}^{c,0}(\bm{q}) \nonumber \\
&+&\frac{1}{2}\left(\!\hat{\Gamma}^{s,0}\!+\!\hat{\Gamma}^{c,0}(\bm{q})\!\right).
\label{eq:pair}
\end{eqnarray}

We perform RPA calculations for each model with $V'=0.464$ eV. 
$V'$ is simply a parameter to adjust the orbital fluctuation strength, which was determined to reproduce the experimental observations at ambient pressure, i.e., to satisfy the condition that the orbital fluctuation is larger than the spin fluctuation. 
Hereafter, we mainly discuss the total spin susceptibility $\chi^{s}(\bm{q},0)=\sum_{l,l'}\chi^{s}_{l,l;l',l'}(\bm{q},0)$ at the Fe site and the orbital susceptibility $\chi^{c}_{x^2-y^2}(\bm{q},0)$. The spin (charge) Stoner factor $\alpha_{s(c)}$ is given by the maximum eigenvalue of $\hat{\Gamma}^{s(c)}(\bm q)\hat{\chi}^0({\bm q},0)$. The magnetic (orbital) order is realized when $\alpha_{s(c)}$ becomes unity. Note that although $\alpha_{s(c)}$ is not experimentally observable, these values are taken as a measure of $T_{m(s)}$.
In the previous study \cite{usui} for the five-orbital model, both $\alpha_s$ and the eigenvalue of the Eliashberg equation were found to monotonically decrease upon applying pressure, in contrast to the experimental observations. The effect of the unusual Fermi surface and the uplift of the Se atomic position were not sufficiently included in Ref.~\citenum{usui}. Also, the pressure dependence of $\alpha_c$ was not clarified.

\subsection{Spin and orbital fluctuations in tetragonal state}
\label{sect3-3}
\begin{figure}[t]
\begin{center}
\includegraphics[width=82mm]{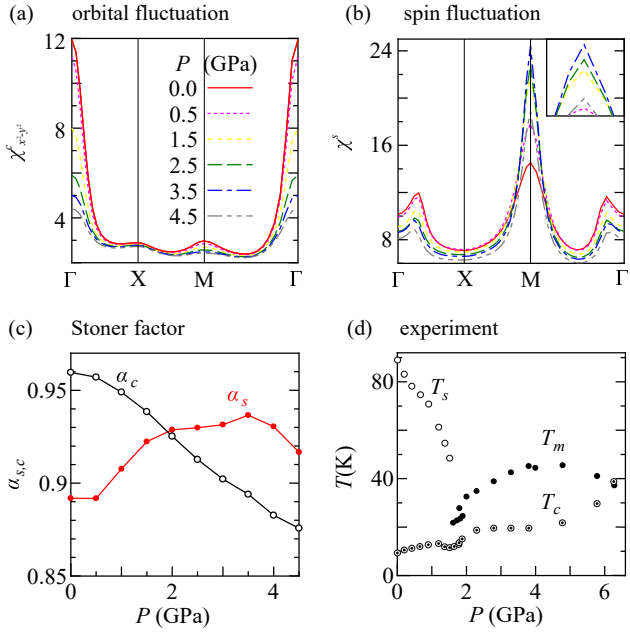}
\caption{(Color online)
 ${\bm q}$ dependence of (a) orbital and (b) spin susceptibilities of FeSe at $P=0.0, 0.5, 1.5, 2.5, 3.5,$ and $4.5$ GPa with $f_d=0.279$. (c) Pressure dependence of charge and spin Stoner factors, which are taken as measures of $T_{s}$ and $T_{m}$, respectively. (d) Experimental result of temperature vs pressure from Ref.~\citenum{sun}.
\label{fig_chi}}
\end{center}
\end{figure}

Figures \ref{fig_chi}(a) and \ref{fig_chi}(b) respectively show the ${\bm q}$ dependences of the orbital and spin susceptibilities for $P=0.0, 0.5, 1.5, 2.5, 3.5,$ and $4.5$ GPa with $T=0.03$ eV, $f_d=0.279$, and $f_p=0.25$. One observes that the orbital susceptibility for ${\bm q}=(0,0)$ is enhanced owing to the $d$-$p$ orbital-polarization interaction at ambient pressure. It decreases as $P$ increases, since the shrinkage of the two-hole Fermi surfaces yields the reduction of $zx/yz$ orbital contribution to the Fermi energy. Conversely, the spin susceptibility for ${\bm q}=(\pi,\pi)$ is strongly enhanced by applying pressure for $P<3.5$ GPa and decreases moderately for $P>3.5$ GPa.

In order to understand this behavior, we study the orbital-resolved spin susceptibility. In the low-pressure region, the total spin fluctuation $\chi^{s}$ is composed of mainly $\chi^{s}_{l,l;l,l}$ of $l=zx/yz$ \cite{note_chis} due to the Fermi surface nesting of the $zx/yz$ orbitals with the wave vector ${\bm q}=(\pi,\pi)$. Note that the spin fluctuation is not so strong owing to the smallness of the Fermi surfaces. In the high-pressure region, the hole Fermi surface, which primarily consists of the $xy$ orbital, supersedes the $zx/yz$ hole Fermi surfaces, and then $\chi^{s}_{l,l;l,l}$ of $l=xy$ is dominant, which originates from the $xy$ intraorbital nesting between the electron and hole Fermi surfaces. In contrast to the spin fluctuation, $\chi^{c}_{x^2-y^2}$ is primarily composed of $\chi^{c}_{l,l;l',l'}$ of $(l,l')=(zx/yz,zx/yz)$ with ${\bm q}=(0,0)$ at all pressures. This component does not require on the electron-hole Fermi surface nesting of the $zx/yz$ orbital with ${\bm q}=(\pi,\pi)$ or the $xy$ orbital contribution to the Fermi surfaces. In other words, the $d$-$p$ orbital-polarization interaction only enhances the ferro-orbital fluctuation owing to the $zx/yz$ orbital contribution to the Fermi surfaces. Note that the charge susceptibility $\sum_{l,l'}\chi^{c}_{l,l;l',l'}$ with ${\bm q}=(0,0)$ is also enhanced owing to the $d$-$p$ Coulomb interaction $V$.

Figure \ref{fig_chi}(c) shows the spin and charge Stoner factors as a function of pressure. Similarly to the spin and orbital susceptibilities, in the low-pressure region up to $3.5$ GPa, $\alpha_s$ markedly increases while $\alpha_c$ is deceased, leading to noteworthy opposite pressure dependences of the spin and orbital fluctuations for the reasons mentioned above. For $P>3.5$ GPa, $\alpha_s$ and $\alpha_c$ are decreased and $\alpha_s$ exhibits a single-dome shape at 3.5 GPa. These results are in good agreement with the experimental phase diagram of $T_{s(m)}$ vs $P$ in Ref.~\citenum{sun} [shown in Fig.~\ref{fig_chi}(d)].

\subsection{Superconductivity in orthorhombic state}
\label{sect3-4}
In this subsection, we address the pressure-induced superconductivity mediated by the pressure-induced spin fluctuation in the orthorhombic phase. We study uniform orbital-ordered and $\bm k$-dependent orbital-ordered states, the latter of which is written as $\delta\Sigma_{zx(yz)}=-(+)\frac{1}{4}(\delta E^{\Gamma}_{\rm nem}+\delta E^{M}_{\rm nem})+(-)\frac{1}{8}(\delta E^{M}_{\rm nem}-\delta E^{\Gamma}_{\rm nem})(\cos k_x+\cos k_y)$, namely a sign-reversing orbital splitting \cite{suzuki,onari_FeSe}. Here, $\delta E_{\rm nem}=E_{zx}-E_{yz}$ is the orbital energy splitting at the $\Gamma$ and M points. We set $\delta E^{\Gamma}_{\rm nem}=-0.05$ eV and $\delta E^{M}_{\rm nem}=0.15$ eV. This orbital order yields a good nesting of the Fermi surface for the $yz$ and $xy$ orbital components but poor nesting for the $zx$ component, so the $yz$ and $xy$ orbital components of the spin fluctuation are enhanced.

\begin{figure}[t]
\begin{center}
\includegraphics[width=82mm]{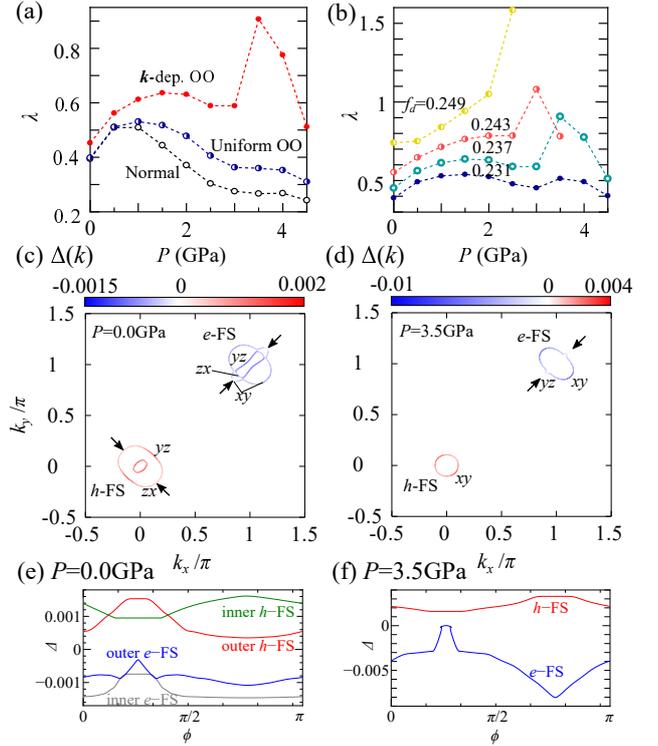}
\caption{(Color online)
 (a) Pressure dependence of the superconducting eigenvalue for $f_d=0.237$ in the tetragonal (Normal), uniform orbital-ordered (Uniform OO), and $\bm k$-dependent orbital-ordered ($\bm k$-dep.~OO) states. (b) Pressure dependence of the superconducting eigenvalue for several $f_d$ values in the $\bm k$-dep.~OO state. (c)-(f) Gap function $\Delta(\bm k,i\pi T)$ on Fermi surfaces at $P=0.0$ and $3.5$ GPa for $f_d=0.237$ in the $\bm k$-dep.~OO state. The black arrows represent the gap minima. 
\label{fig_lambda_chap4}}
\end{center}
\end{figure}

Figures~\ref{fig_lambda_chap4}(a) and \ref{fig_lambda_chap4}(b) show the pressure dependence of the eigenvalue $\lambda$ of the Eliashberg equation for $f_d=0.237$, $V'=0$, and $T=0.01$ eV. $\lambda$ exhibits broad enhancement with a double-dome shape in the $\bm k$-dependent orbital-ordered state. The sharp peak at $P=3.5$ GPa is due to the proximity to the antiferromagnetic order. Indeed, the peak shifts to a low pressure as $f_d$ increases [see Fig.~\ref{fig_lambda_chap4}(b)]. The nonmonotonic increase such as that of the double-dome $\lambda$ is characterized by the switch of the dominant orbital components of the spin fluctuation.
In Figs.~\ref{fig_lambda_chap4}(c)-\ref{fig_lambda_chap4}(f), the gap function shows a large anisotropy, namely nodal-like $s_{\pm}$-wave pairing, which has 	gap minima (represented by arrows in the figures) on the hole and/or electron Fermi surfaces due to the orbital dependence of the spin fluctuation. At $P=0$ GPa, the magnitude of the gap is weak on the Fermi surface composed of the $zx$ orbital, in agreement with experiments \cite{sprau,xu_2}, and it is also weak on the $xy$ orbital.

The obtained result for the $\bm k$-dependent orbital-ordered state seems to be consistent with the experimental phase diagram in the low-pressure region, where $T_c$ increases with increasing pressure. For the high-pressure region, however, our calculation is restricted to the case without the antiferromagnetic order, in contrast to the experiment, and thus we need further investigation including the antiferromagnetic order as well as a full DMFT calculation over the whole pressure regime.

%

\section{Summary and Discussion}

We studied the electronic state of FeSe at ambient pressure and under pressure. 

In Sect.~\ref{sect2}, we presented a systematic analysis of a 16-band $d$-$p$ model of FeSe at ambient pressure within dynamical mean-field theory. We found that the dissipation of a shallow hole pocket occurs owing to the orbital dependence of on-site Coulomb interaction, in agreement with the orbital-dependent band lifting observed by the recent ARPES and QO experiments \cite{maletz,watson}. We also found that the intersite orbital-polarization Coulomb interaction between the Fe $d$ orbital and Se $p$ orbital drives an electric orbital order in the absence of a low-energy commensurate spin response.

In Sect.~\ref{sect3}, we investigated the spin-charge-orbital fluctuation and superconductivity under pressure in the 16-band $d$-$p$ Hubbard model for FeSe in the paramagnetic phase by using the random phase approximation with respect to the Fe 3$d$ Coulomb interaction and the orbital-polarization interaction. The Fermi surfaces obtained from the effective tight-binding model clearly indicate the signature of the topological transition with increasing pressure. We clarified that the ferro-orbital order is monotonically suppressed with increasing pressure, whereas the antiferromagnetic order is robustly realized with a dome shape at 3.5 GPa, in good agreement with experiments. The nodal-like $s_{\pm}$ pairing is realized owing to the orbital dependence of the spin fluctuation in $C_2$ symmetry. The eigenvalue $\lambda$ shows the double-dome shape due to the pressure-induced spin fluctuation.

The effects of several parameters were investigated such as the reduction factors with varying values and the additional hoppings. In the analysis of the pressure dependence, the additional hopping were introduced and justified by both the experimental observation and the DMFT analysis in Sect.~\ref{sect2}. The additional hoppings depress (raise) the orbital energy level at the $\Gamma$- (M-) point.
To obtain insight about the parameter-free pressure dependence, a DMFT calculation at all pressure values will be required and will be explicitly discussed in a subsequent paper.

\section*{Acknowledgments}
This work was partially supported by a Grant-in-Aid for Scientific Research from the Ministry of Education, Culture, Sports, Science and Technology and by JSPS KAKENHI Grant No. 16J01929. J. I. was supported by a JSPS Fellowship for Young Scientists. 

\bibliographystyle{jpsj.bst}


\begin{thebibliography}{10}

\bibitem{kamihara}
Y.~Kamihara, H.~Hiramatsu, M.~Hirano, R.~Kawamura, H.~Yanagi, T.~Kamiya, and
  H.~Hosono, J. Am. Chem. Soc. {\bfseries 128}, 10012 (2006).

\bibitem{mazin}
I.~I. Mazin, D.~J. Singh, M.~D. Johannes, and M.~H. Du, Phys. Rev. Lett.
  {\bfseries 101}, 057003 (2008).

\bibitem{kuroki}
K.~Kuroki, S.~Onari, R.~Arita, H.~Usui, Y.~Tanaka, H.~Kontani, and H.~Aoki,
  Phys. Rev. Lett. {\bfseries 101}, 087004 (2008).

\bibitem{ikeda}
H.~Ikeda, J. Phys. Soc. Jpn. {\bfseries 77}, 123707 (2008).

\bibitem{thomale}
R.~Thomale, C.~Platt, W.~Hanke, and B.~A. Bernevig, Phys. Rev. Lett. {\bfseries
  106}, 187003 (2011).

\bibitem{wang}
X.~C. Wang, Q.~Q. Liu, Y.~X. Lv, W.~B. Gao, L.~X. Yang, R.~C. Yu, F.~Y. Li, and
  C.~Q. Jin, Solid State Commun. {\bfseries 148}, 538 (2008).

\bibitem{kontani_orbiton}
H.~Kontani, T.~Saito, and S.~Onari, Phys. Rev. B {\bfseries 84}, 024528 (2011).

\bibitem{yanagi_e-ph_2}
Y.~Yanagi, Y.~Yamakawa, N.~Adachi, and Y.~\=Ono, J. Phys. Soc. Jpn. {\bfseries
  79}, 123707 (2010).

\bibitem{fernandes}
R.~M. Fernandes, L.~H. VanBebber, S.~Bhattacharya, P.~Chandra, V.~Keppens,
  D.~Mandrus, M.~A. McGuire, B.~C. Sales, A.~S. Sefat, and J.~Schmalian, Phys.
  Rev. Lett. {\bfseries 105}, 157003 (2010).

\bibitem{fernandes_2}
R.~M. Fernandes, A.~V. Chubukov, and J.~Schmalian, Nat. Mater. {\bfseries 10},
  97 (2014).

\bibitem{kruger}
F.~Kr\"uger, S.~Kumar, J.~Zaanen, and J.~van~den Brink, Phys. Rev. B {\bfseries
  79}, 054504 (2009).

\bibitem{onari_sc-vc}
S.~Onari and H.~Kontani, Phys. Rev. Lett. {\bfseries 109}, 137001 (2012).

\bibitem{yamada_upd}
T.~Yamada, J.~Ishizuka, and Y.~\=Ono, J. Phys. Soc. Jpn. {\bfseries 83}, 043704
  (2014).

\bibitem{mizuguchi_3}
Y.~Mizuguchi, F.~Tomioka, S.~Tsuda, T.~Yamaguchi, and Y.~Takano, J. Phys. Soc.
  Jpn. {\bfseries 78}, 074712 (2009).

\bibitem{watson_2}
M.~D. Watson, T.~K. Kim, A.~A. Haghighirad, S.~F. Blake, N.~R. Davies,
  M.~Hoesch, T.~Wolf, and A.~I. Coldea, Phys. Rev. B {\bfseries 92}, 121108
  (2015).

\bibitem{kawasaki}
Y.~Kawasaki, K.~Deguchi, S.~Demura, T.~Watanabe, H.~Okazaki, T.~Ozaki,
  T.~Yamaguchi, H.~Takeya, and Y.~Takano, Solid State Commun. {\bfseries 152},
  1135  (2012).

\bibitem{medvedev}
S.~Medvedev, T.~M. McQueen, I.~Trojan, T.~Palasyuk, M.~I. Eremets, R.~J. Cava,
  S.~Naghavi, F.~Casper, V.~Ksenofontov, G.~Wortmann, and C.~Felser, Nat.
  Mater. {\bfseries 8}, 630 (2009).

\bibitem{urata}
T.~Urata, Y.~Tanabe, K.~K. Huynh, Y.~Yamakawa, H.~Kontani, and K.~Tanigaki,
  Phys. Rev. B {\bfseries 93}, 014507 (2016).

\bibitem{zvyagina}
{Zvyagina, G. A.}, {Gaydamak, T. N.}, {Zhekov, K. R.}, {Bilich, I. V.}, {Fil,
  V. D.}, {Chareev, D. A.}, and {Vasiliev, A. N.}, EPL {\bfseries 101}, 56005
  (2013).

\bibitem{massat}
P.~Massat, D.~Farina, I.~Paul, S.~Karlsson, P.~Strobel, P.~Toulemonde, M.-A.
  M\'easson, M.~Cazayous, A.~Sacuto, S.~Kasahara, T.~Shibauchi, Y.~Matsuda, and
  Y.~Gallais, Proceedings of the National Academy of Sciences {\bfseries 113},
  9177 (2016).

\bibitem{goto}
T.~Goto, R.~Kurihara, K.~Araki, K.~Mitsumoto, M.~Akatsu, Y.~Nemoto,
  S.~Tatematsu, and M.~Sato, J. Phys. Soc. Jpn. {\bfseries 80}, 073702 (2011).

\bibitem{yoshizawa}
M.~Yoshizawa, D.~Kimura, T.~Chiba, A.~Ismayil, Y.~Nakanishi, K.~Kihou, C.~Lee,
  A.~Iyo, H.~Eisaki, M.~Nakajima, and S.~Uchida, J. Phys. Soc. Jpn. {\bfseries
  81}, 024604 (2012).

\bibitem{gallais}
Y.~Gallais, R.~M. Fernandes, I.~Paul, L.~Chauvi\`ere, Y.-X. Yang, M.-A.
  M\'easson, M.~Cazayous, A.~Sacuto, D.~Colson, and A.~Forget, Phys. Rev. Lett.
  {\bfseries 111}, 267001 (2013).

\bibitem{baek_3}
S.-H. Baek, D.~V. Efremov, J.~M. Ok, J.~S. Kim, J.~van~den Brink, and
  B.~B\"uchner, Nat. Mater. {\bfseries 14}, 210 (2015).

\bibitem{bohmer}
A.~E. B\"ohmer, T.~Arai, F.~Hardy, T.~Hattori, T.~Iye, T.~Wolf, H.~v.
  L\"ohneysen, K.~Ishida, and C.~Meingast, Phys. Rev. Lett. {\bfseries 114},
  027001 (2015).

\bibitem{mukherjee}
S.~Mukherjee, A.~Kreisel, P.~J. Hirschfeld, and B.~M. Andersen, Phys. Rev.
  Lett. {\bfseries 115}, 026402 (2015).

\bibitem{maletz}
J.~Maletz, V.~B. Zabolotnyy, D.~V. Evtushinsky, S.~Thirupathaiah, A.~U.~B.
  Wolter, L.~Harnagea, A.~N. Yaresko, A.~N. Vasiliev, D.~A. Chareev, A.~E.
  B\"ohmer, F.~Hardy, T.~Wolf, C.~Meingast, E.~D.~L. Rienks, B.~B\"uchner, and
  S.~V. Borisenko, Phys. Rev. B {\bfseries 89}, 220506 (2014).

\bibitem{watson}
M.~D. Watson, T.~K. Kim, A.~A. Haghighirad, N.~R. Davies, A.~McCollam,
  A.~Narayanan, S.~F. Blake, Y.~L. Chen, S.~Ghannadzadeh, A.~J. Schofield,
  M.~Hoesch, C.~Meingast, T.~Wolf, and A.~I. Coldea, Phys. Rev. B {\bfseries
  91}, 155106 (2015).

\bibitem{shimojima_str_2}
T.~Shimojima, Y.~Suzuki, T.~Sonobe, A.~Nakamura, M.~Sakano, J.~Omachi,
  K.~Yoshioka, M.~Kuwata-Gonokami, K.~Ono, H.~Kumigashira, A.~E. B\"ohmer,
  F.~Hardy, T.~Wolf, C.~Meingast, H.~v. L\"ohneysen, H.~Ikeda, and K.~Ishizaka,
  Phys. Rev. B {\bfseries 90}, 121111 (2014).

\bibitem{suzuki}
Y.~Suzuki, T.~Shimojima, T.~Sonobe, A.~Nakamura, M.~Sakano, H.~Tsuji,
  J.~Omachi, K.~Yoshioka, M.~Kuwata-Gonokami, T.~Watashige, R.~Kobayashi,
  S.~Kasahara, T.~Shibauchi, Y.~Matsuda, Y.~Yamakawa, H.~Kontani, and
  K.~Ishizaka, Phys. Rev. B {\bfseries 92}, 205117 (2015).

\bibitem{onari_FeSe}
S.~Onari, Y.~Yamakawa, and H.~Kontani, Phys. Rev. Lett. {\bfseries 116}, 227001
  (2016).

\bibitem{yamakawa_FeSe_sc-vc}
Y.~Yamakawa, S.~Onari, and H.~Kontani, Phys. Rev. X {\bfseries 6}, 021032
  (2016).

\bibitem{yamakawa_FeSe_p}
Y.~Yamakawa and H.~Kontani, Phys. Rev. B {\bfseries 96}, 144509 (2017).

\bibitem{imai}
T.~Imai, K.~Ahilan, F.~L. Ning, T.~M. McQueen, and R.~J. Cava, Phys. Rev. Lett.
  {\bfseries 102}, 177005 (2009).

\bibitem{bendele}
M.~Bendele, A.~Amato, K.~Conder, M.~Elender, H.~Keller, H.-H. Klauss,
  H.~Luetkens, E.~Pomjakushina, A.~Raselli, and R.~Khasanov, Phys. Rev. Lett.
  {\bfseries 104}, 087003 (2010).

\bibitem{bendele_2}
M.~Bendele, A.~Ichsanow, Y.~Pashkevich, L.~Keller, T.~Str\"assle, A.~Gusev,
  E.~Pomjakushina, K.~Conder, R.~Khasanov, and H.~Keller, Phys. Rev. B
  {\bfseries 85}, 064517 (2012).

\bibitem{terashima}
T.~Terashima, N.~Kikugawa, S.~Kasahara, T.~Watashige, T.~Shibauchi, Y.~Matsuda,
  T.~Wolf, A.~E. B\"ohmer, F.~Hardy, C.~Meingast, H.~v.~L\"ohneysen, and
  S.~Uji, J. Phys. Soc. Jpn. {\bfseries 84}, 063701 (2015).

\bibitem{sun}
J.~P. Sun, K.~Matsuura, G.~Z. Ye, Y.~Mizukami, M.~Shimozawa, K.~Matsubayashi,
  M.~Yamashita, T.~Watashige, S.~Kasahara, Y.~Matsuda, J.-Q. Yan, B.~C. Sales,
  Y.~Uwatoko, J.~G. Cheng, and T.~Shibauchi, Nat. Commun. {\bfseries 7}, 12146
  (2016).

\bibitem{miyoshi_2}
K.~Miyoshi, K.~Morishita, E.~Mutou, M.~Kondo, O.~Seida, K.~Fujiwara,
  J.~Takeuchi, and S.~Nishigori, J. Phys. Soc. Jpn. {\bfseries 83}, 013702
  (2014).

\bibitem{kothapalli}
K.~Kothapalli, A.~E. B\"ohmer, W.~T. Jayasekara, B.~G. Ueland, P.~Das,
  A.~Sapkota, V.~Taufour, Y.~Xiao, E.~E. Alp, S.~L. Bud'ko, P.~C. Canfield,
  A.~Kreyssig, and A.~I. Goldman, Nat. Commun. {\bfseries 7}, 12728 (2016).

\bibitem{terashima_2}
T.~Terashima, N.~Kikugawa, A.~Kiswandhi, D.~Graf, E.-S. Choi, J.~S. Brooks,
  S.~Kasahara, T.~Watashige, Y.~Matsuda, T.~Shibauchi, T.~Wolf, A.~E. B\"ohmer,
  F.~Hardy, C.~Meingast, H.~v. L\"ohneysen, and S.~Uji, Phys. Rev. B {\bfseries
  93}, 094505 (2016).

\bibitem{wang_6}
P.~S. Wang, S.~S. Sun, Y.~Cui, W.~H. Song, T.~R. Li, R.~Yu, H.~Lei, and W.~Yu,
  Phys. Rev. Lett. {\bfseries 117}, 237001 (2016).

\bibitem{liebsch_2}
A.~Liebsch and H.~Ishida, Phys. Rev. B {\bfseries 82}, 155106 (2010).

\bibitem{aichhorn_3}
M.~Aichhorn, S.~Biermann, T.~Miyake, A.~Georges, and M.~Imada, Phys. Rev. B
  {\bfseries 82}, 064504 (2010).

\bibitem{yin_power-low}
Z.~P. Yin, K.~Haule, and G.~Kotliar, Phys. Rev. B {\bfseries 86}, 195141
  (2012).

\bibitem{miyake}
T.~Miyake, K.~Nakamura, R.~Arita, and M.~Imada, J. Phys. Soc. Jpn. {\bfseries
  79}, 0447105 (2010).

\bibitem{hirayama}
M.~Hirayama, T.~Miyake, and M.~Imada, Phys. Rev. B {\bfseries 87}, 195144
  (2013).

\bibitem{scherer}
D.~D. Scherer, A.~C. Jacko, C.~Friedrich, E.~\ifmmode \mbox{\c{S}}\else
  \c{S}\fi{}a\ifmmode \mbox{\c{s}}\else \c{s}\fi{}\ifmmode \imath \else \i
  \fi{}o\ifmmode~\breve{g}\else \u{g}\fi{}lu, S.~Bl\"ugel, R.~Valent\'{\i}, and
  B.~M. Andersen, Phys. Rev. B {\bfseries 95}, 094504 (2017).

\bibitem{ishizuka_s+-s++}
J.~Ishizuka, T.~Yamada, Y.~Yanagi, and Y.~\=Ono, J. Phys. Soc. Jpn. {\bfseries
  82}, 123712 (2013).

\bibitem{ishizuka_hole-s+-}
J.~Ishizuka, T.~Yamada, Y.~Yanagi, and Y.~\=Ono, Journal of the Physical
  Society of Japan {\bfseries 85}, 114709 (2016).

\bibitem{marzari}
N.~Marzari and D.~Vanderbilt, Phys. Rev. B {\bfseries 56}, 12847 (1997).

\bibitem{souza}
I.~Souza, N.~Marzari, and D.~Vanderbilt, Phys. Rev. B {\bfseries 65}, 035109
  (2001).

\bibitem{mostofi}
A.~A. Mostofi, J.~R. Yates, Y.-S. Lee, I.~Souza, D.~Vanderbilt, and N.~Marzari,
  Computer Physics Communications {\bfseries 178}, 685  (2008).

\bibitem{kunes}
J.~Kune\v{s}, R.~Arita, P.~Wissgott, A.~Toschi, H.~Ikeda, and K.~Held, Computer
  Physics Communications {\bfseries 181}, 1888  (2010).

\bibitem{blaha_2}
P. Blaha, K. Schwarz, G. Madsen, D. Kvasnicka, and J. Luitz, WIEN2k, An
  Augmented Plane Wave + Local Orbitals Program for Calculating Crystal
  Properties (Tech. Univ. Wien, Vienna, 2001).

\bibitem{wien2k}
We used the full-potential linearized augmented plane-wave method within the
  generalized gradient approximation as implemented in the WIEN2k code.
  \cite{blaha_2} The maximum reciprocal lattice vector $K_{max}$ was chosen as
  $R_{\rm MT}K_{\rm max}=7.0$ and 25$\times$25$\times$15 $k$-points sampling
  was used for the self-consistent calculation.

\bibitem{park}
H.~Park, K.~Haule, and G.~Kotliar, Phys. Rev. Lett. {\bfseries 107}, 137007
  (2011).

\bibitem{anisimov_fll}
V.~I. Anisimov, F.~Aryasetiawan, and A.~I. Lichtenstein, Journal of Physics:
  Condensed Matter {\bfseries 9}, 767 (1997).

\bibitem{margadonna}
S.~Margadonna, Y.~Takabayashi, Y.~Ohishi, Y.~Mizuguchi, Y.~Takano, T.~Kagayama,
  T.~Nakagawa, M.~Takata, and K.~Prassides, Phys. Rev. B {\bfseries 80}, 064506
  (2009).

\bibitem{kuroki_4}
K.~Kuroki, Solid State Commun. {\bfseries 152}, 711  (2012).

\bibitem{usui}
H.~Usui and K.~Kuroki, Physica C: Superconductivity {\bfseries 470}, Supplement
  1, S382  (2010).

\bibitem{note_chis}
At the present temperature $T=30$\,meV, the $xy$ orbital component of the spin
  susceptibility is larger than that of the $zx$ orbital because of the
  comparable energy scale of $E_{\Gamma}=50$\,meV. When we assume realistic low
  temperatures, the $zx$ orbital component becomes larger than that the $xy$
  orbital. The pressure dependence of the Stoner factor is qualitatively not
  changed in this case in low temperatures.

\bibitem{sprau}
P.~O. Sprau, A.~Kostin, A.~Kreisel, A.~E. B{\"o}hmer, V.~Taufour, P.~C.
  Canfield, S.~Mukherjee, P.~J. Hirschfeld, B.~M. Andersen, and J.~C.~S. Davis,
  Science {\bfseries 357}, 75 (2017).

\bibitem{xu_2}
H.~C. Xu, X.~H. Niu, D.~F. Xu, J.~Jiang, Q.~Yao, Q.~Y. Chen, Q.~Song,
  M.~Abdel-Hafiez, D.~A. Chareev, A.~N. Vasiliev, Q.~S. Wang, H.~L. Wo,
  J.~Zhao, R.~Peng, and D.~L. Feng, Phys. Rev. Lett. {\bfseries 117}, 157003
  (2016).

\end{thebibliography}
\end{document}